\documentclass[11pt]{article}

\newif\ifarxiv
\arxivtrue

\ifarxiv
    \usepackage[preprint]{acl}
\else
    \usepackage[review]{acl}
\fi

\usepackage{times}
\usepackage{latexsym}
\usepackage[T1]{fontenc}
\usepackage[utf8]{inputenc}
\usepackage{microtype}
\usepackage{inconsolata}
\usepackage{amsmath}
\usepackage{amssymb}
\usepackage{booktabs}
\usepackage{graphicx}
\usepackage{xspace}
\usepackage{enumitem}
\usepackage{xcolor}
\usepackage{tikz}
\usepackage{url}
\usetikzlibrary{shapes.geometric,arrows.meta,positioning}

\newcommand{\ndcg}{\ensuremath{\mathrm{nDCG@10}}\xspace}
\newcommand{\best}[1]{\textbf{#1}}
\newcommand{\second}[1]{\underline{#1}}

\definecolor{storeblue}{RGB}{210,228,245}
\definecolor{storeblueline}{RGB}{38,97,156}
\definecolor{modelpurple}{RGB}{226,214,243}
\definecolor{modelpurpleline}{RGB}{121,76,178}
\definecolor{decideorange}{RGB}{255,233,200}
\definecolor{decideorangeline}{RGB}{214,137,16}
\definecolor{transorange}{RGB}{255,224,178}
\definecolor{transorangeline}{RGB}{198,108,12}
\definecolor{artgreen}{RGB}{212,237,213}
\definecolor{artgreenline}{RGB}{56,142,60}
\tikzset{
  store/.style   ={cylinder, shape border rotate=90, aspect=0.22, draw=storeblueline,
                   fill=storeblue, text width=2.0cm, align=center, minimum height=0.9cm, font=\small},
  model/.style   ={trapezium, trapezium left angle=70, trapezium right angle=110, draw=modelpurpleline,
                   fill=modelpurple, text width=2.2cm, align=center, minimum height=0.85cm, font=\small},
  decide/.style  ={diamond, aspect=1.8, draw=decideorangeline, fill=decideorange,
                   text width=1.8cm, align=center, inner sep=1pt, font=\small},
  trans/.style   ={rectangle, rounded corners=1pt, draw=transorangeline, fill=transorange,
                   text width=2.2cm, align=center, minimum height=0.85cm, font=\small},
  artifact/.style={rectangle, draw=artgreenline, fill=artgreen,
                   text width=2.2cm, align=center, minimum height=0.85cm, font=\small},
  eval/.style    ={ellipse, draw=artgreenline, fill=artgreen,
                   text width=1.9cm, align=center, minimum height=0.85cm, font=\small},
  flow/.style    ={-{Latex[length=2.2mm]}, thick},
  prov/.style    ={-{Latex[length=2.0mm]}, dashed, storeblueline, thick},
  phaselab/.style={font=\footnotesize\bfseries, text=black!60, anchor=west},
}

\title{PETRA: Transforming Web Text for Petroleum-Engineering Domain Adaptation}

\ifarxiv

    \author{%
    \begin{tabular}{c}
    Kirill Dubovikov$^{1}$,
    Omar El Mansouri$^{1}$,
    Hachem Madmoun$^{1}$,
    Yanda Li$^{1}$,
    Sandeep Kumar$^{1}$,
    Aya El Mir$^{1}$, \\
    Supriyo Ghosh$^{2}$,
    Writabrata Bhattacharya$^{2}$,
    Adrian Garcia-Garcia$^{2}$,
    Onkar Pandit$^{2}$,
    Sunil Kumar Sahu$^{2}$, \\
    Federico Castanedo$^{2}$,
    Larry Murray$^{2}$,
    Martin Tak\'a\v{c}$^{1}$,
    Salem Lahlou$^{1}$ \\[0.5em]
    $^{1}$Mohamed bin Zayed University of Artificial Intelligence \\
    $^{2}$Inception AI \\[0.25em]
    \texttt{Kirill.Dubovikov@mbzuai.ac.ae}
    \end{tabular}
    }

\else
    \author{Anonymous EMNLP 2026 Industry Track Submission}
\fi

\begin{document}
\maketitle

\begin{abstract}
    Petroleum-engineering search exposes a supervision gap for strong general retrievers: relevant evidence exists in public web text, but domain relevance labels are scarce. To address this gap, we propose PETRA, a large-scale Petroleum Engineering Text for Retrieval Adaptation dataset and pipeline that converts noisy public web data into a curated domain corpus and synthetic supervision for dense retrieval and reranking. PETRA contains 1.36M curated chunks, approximately 2B token equivalents, $\approx$859k, embedding training rows from $\approx$224k anchors, and roughly 400k teacher-scored reranker candidate rows. Its construction combines high-recall energy-domain curation, an energy-domain classifier with 98.4\% test accuracy, chunk-grounded query generation, LLM-written hard negatives, and retrieval-mined candidate lists. PETRA improves first-stage in-domain Normalized Discounted Cumulative Gain (nDCG) from 0.703 to 0.763 through score fusion. Reranker adaptation improves the public Earth Science benchmark by 44\% relative and a six-task reasoning-intensive panel by 23\%. Failed training recipes show that high train-holdout accuracy on synthetic labels does not predict retrieval gains; retrieval-mined data helps only after being repackaged as teacher-scored candidate lists sampled from the inference-time candidate distribution. We release an anonymized version of PETRA at: ~\url{https://huggingface.co/datasets/petra-2026/PETRA}
\end{abstract}

\section{Introduction}
\label{sec:introduction}

Energy production is a knowledge-intensive industry in which operational decisions depend on large amounts of technical, procedural, and scientific evidence. In practice, this evidence is scattered across standard operating procedures, equipment manuals, engineering studies, incident reports, and scientific literature. Retrieval-augmented systems \citep{DBLP:journals/corr/abs-2108-07258,DBLP:conf/emnlp/0044CJWJW024,qwen2025embedding} are attractive in this setting because they can ground answers in source documents. However, this shifts the burden to retrieval: in petroleum engineering, relevance often depends on domain conventions that are not captured by surface lexical overlap. Acronyms such as EUR, IP30, and OOIP, polysemous terms such as \emph{well}, \emph{field}, \emph{reservoir}, \emph{play}, and \emph{spread}, equipment names, unit conventions, and procedural context can all determine whether a passage is actually relevant. Such mismatches can cause retrievers to favor topically related passages over passages that contain the required procedural or numerical evidence.

Existing data sources do not resolve the supervision gap for petroleum-engineering retrieval adaptation. Curated domain datasets and expert-written corpora \citep{gunasekar2023textbooks,lima} are cleaner, but they are usually too small or too narrow to cover the terminology, procedures, and equipment contexts needed by a deployed retriever. Large web corpora \citep{li2024dclm,penedo2024fineweb} provide scale, but contain off-domain pages, duplicated text, OCR artifacts, weak structure, and uneven metadata; they also do not provide query-passage relevance labels. Public retrieval benchmarks \citep{thakur2021beir,su2025bright} help measure general retrieval quality, but they do not provide petroleum-engineering supervision for adapting a production search stack.

We introduce PETRA, a Petroleum Engineering Text for Retrieval Adaptation dataset and pipeline for converting noisy public web data into domain retrieval supervision. PETRA pairs a curated petroleum-engineering corpus with synthetic query-passage training data for both first-stage retrieval and reranking. The pipeline filters and validates open-source documents, then generates chunk-grounded queries, LLM-written hard negatives, and retrieval-mined candidate lists for adapting a two-stage retrieval stack. We evaluate PETRA in a retrieval system, measuring both in-domain gains and out-of-domain retention, and are currently working with enterprise Oil and Gas customers to validate the performance results under different use-cases.

To the best of our knowledge, PETRA is the first publicly described large-scale dataset and pipeline for petroleum-engineering retrieval adaptation, pairing a curated English petroleum corpus with synthetic supervision for both dense retrieval and reranking.  Our main contributions are as follows:
\begin{itemize}[itemsep=1pt,topsep=2pt,leftmargin=1.2em]
\item We introduce a scalable pipeline for transforming noisy public web text into large-scale petroleum-engineering retrieval data. The pipeline combines high-recall corpus curation, an energy-domain classifier achieving 98.4\% test accuracy, and synthetic supervision for both retrieval and reranking.
\item We present PETRA, a large-scale energy-domain retrieval dataset comprising 1.36M chunks, approximately 2B token equivalents, 859,841 embedding-training rows derived from 224,920 anchors, and roughly 400K teacher-scored reranker candidate rows. We release the public-source portion of the dataset where licensing permits.
\item We conduct an adaptation study of a two-stage retrieval stack, quantifying the tradeoff between domain specialization and general-domain retention. Score fusion improves first-stage in-domain retrieval from 0.703 to 0.763 $\mathrm{nDCG}$, while reranker adaptation yields relative gains of 44\% on Earth Science and 23\% on a six-task reasoning panel.
\item We report negative results and lessons learned, showing that high generated-label accuracy did not guarantee retrieval quality until reranker training matched the inference-time candidate distribution.
\end{itemize}

\section{Related Work}
\label{sec:related}

Our curation design follows evidence that smaller, higher-quality corpora can outperform larger noisy ones \citep{gunasekar2023textbooks,lima,li2024dclm,penedo2024fineweb}. We apply this principle to petroleum engineering, where resources such as K2, GeoGalactica, EnergyGPT, and PetroNLP \citep{deng2024k2,lin2024geogalactica,chebbi2025energygpt,cordeiro2024petronlp} are mainly broad-domain pretraining corpora or non-English task resources rather than refined English retrieval corpora with relevance supervision. Concurrent work uses the same curated corpus for synthetic QA generation and post-training, while PETRA focuses on retrieval and reranking supervision~\citep{anon_overlap_2026}.

PETRA also builds on work in dense retrieval, reranking, synthetic query generation, and hard-negative mining. Strong embedding models \citep{li2023gte,xiao2024cpack,wang-etal-2024-improving-text,lee2024gecko}, MTEB-style evaluation \citep{muennighoff-etal-2023-mteb}, cross-encoder rerankers \citep{nogueira-etal-2020-document,zhuang2023rankt5}, and listwise LLM rankers \citep{sun-etal-2023-chatgpt,zhuang-etal-2024-beyond} provide the retrieval setting for our two-stage stack. For supervision, Promptagator, GPL, InPars, and UDAPDR generate synthetic queries, mine negatives, or distill relevance signals \citep{dai2022promptagator,wang2022gpl,bonifacio2022inpars,saad-falcon-etal-2023-udapdr}, while ANCE and RocketQA show the value of mined hard negatives and cross-encoder denoising \citep{xiong2021ance,qu2021rocketqa}.

Our pipeline differs in how this supervision is packaged. LLM-written negatives are validated against chunk-grounded queries, while reranker examples are created from candidate lists mined from the deployed RAG stack and re-scored with a teacher model, matching the inference-time candidate distribution. This design is motivated by prior work on hard-negative quality, contrastive learning, distillation, domain-adaptive training, and adapter merging
\citep{thakur2025hardnegatives,oord2018cpc,karpukhin2020dpr,wang2022e5,hinton2015distilling,gururangan-etal-2020-dont,yadav2023ties,yu2024dare,wortsman2022modelsoups,ilharco2023task}. Closest to our application are domain-specific retrieval and industrial NLP systems spanning finance, telecom, science, long-context retrieval, industrial documents, asset operations, and oil-and-gas annotation \citep{anderson-etal-2024-greenback,ethiraj-etal-2025-vec,bhattacharjee-etal-2024-indus,zhang-etal-2024-mgte,choi-etal-2024-rradistill,lim-etal-2025-distilling,constantinides-etal-2025-generalized,correia-etal-2025-analysis}.

\section{The PETRA Dataset}
\label{sec:dataset}

\begin{figure*}[t]
  \centering
  \includegraphics[width=\textwidth]{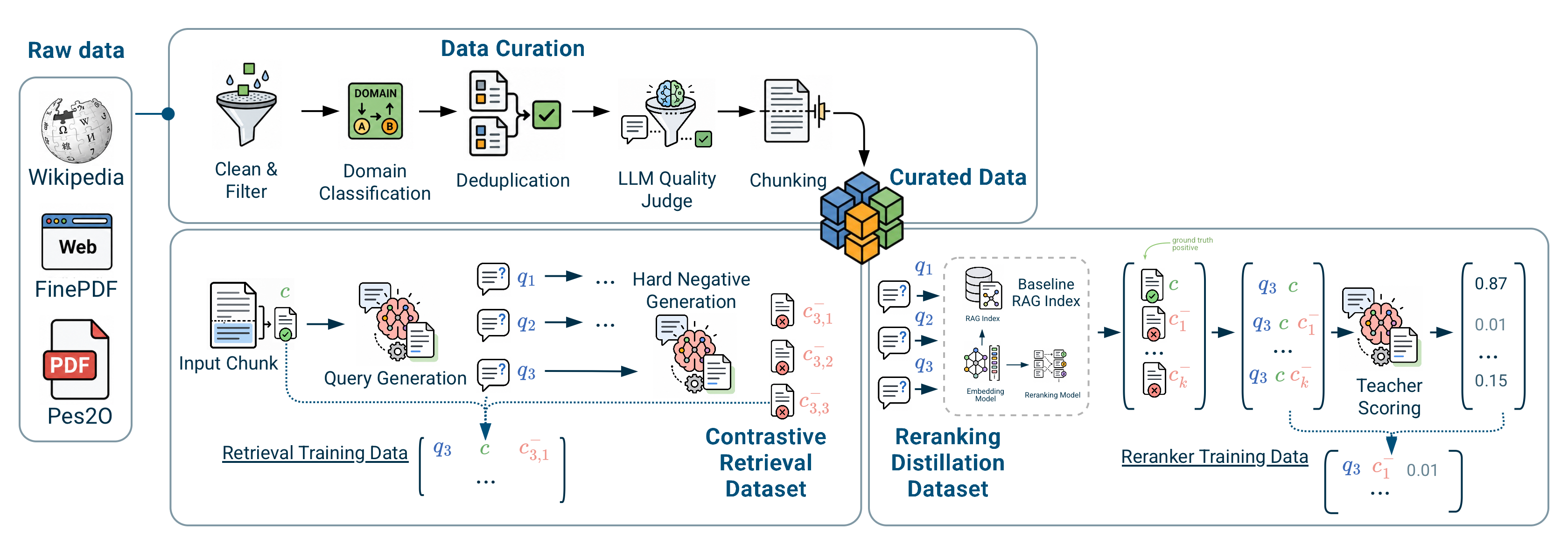}
  \caption{PETRA data construction pipeline. Curation distills open sources into the curated
corpus (\S\ref{subsec:curation}); synthetic supervision converts those chunks
into training data (\S\ref{subsec:supervision})}
  \label{fig:pipeline}
\vspace{-3mm}
\end{figure*}

PETRA turns noisy public web text into retrieval supervision through two stages, shown in Figure~\ref{fig:pipeline}. The first stage curates a high-recall petroleum-engineering corpus, then removes off-domain, duplicated, low-quality, and weakly answerable chunks (\S\ref{subsec:curation}). The second stage converts retained chunks into the training surfaces used by our two-stage retrieval stack (\S\ref{subsec:supervision}): contrastive triples for first-stage retrieval and teacher-scored candidate rows for reranking. Retrieval supervision is anchored in generated query styles observed in user testing, while reranker supervision is mined from baseline candidate lists so that training examples reflect the distribution scored at inference time.

\subsection{Corpus Curation}
\label{subsec:curation}

The curation stage removes off-domain, duplicated, and low-quality text. We aim to achieve a high in-domain recall first, and then optimize for precision in later stages. We implement curation as a resumable Ray pipeline \citep{ray} built on NeMo Curator \citep{nemo_curator} and process three public source families: FinePDFs-Edu, peS2o scientific papers, and a petroleum-engineering slice of English Wikipedia. These sources provide broad coverage while still requiring domain-specific filtering at retrieval-training scale. Each record retains its text, stable identifier, and source metadata, including title, URL, source family, and license.

\paragraph{Source Curation.}

Public web documents contain redundant text and OCR artifacts that add noise to domain adaptation. We remove references, inline citations, page markers and DOIs before applying document-level filters. We then apply document-level heuristics
(Appendix~\ref{app:curation}) to remove the bulk of unusable text by word
count, non-alphanumeric ratio, repeated-line and repeated-paragraph
fractions, each recording its decision signal. A fastText language filter
\citep{fasttext,fasttext_langid} retains English text without discarding short, symbol-heavy technical passages.

To remove remaining off-domain documents, we apply a LoRA-tuned \citep{lora} Llama-3.1-8B \citep{llama3} binary energy classifier trained on labels distilled from Mistral-Large-3-675B \citep{mistral}.
Our classifier serves as a high-recall energy gate, reaching 98.39\%
test accuracy and 99.69\% recall (Appendix~\ref{app:curation}). High recall is crucial for this filter, as a false negative silently discards potentially high-quality in-domain
evidence. Remaining documents are also scored by NeMo Curator quality and safety classifiers.

\paragraph{Chunk Curation.}
\label{sec:chunk}
We reduce duplication before chunking with two-stage deduplication. Exact deduplication removes MD5-identical documents; semantic deduplication
embeds documents with all-MiniLM-L6-v2 \citep{reimers2019sentencebert},
clusters them with $k$-means, and keeps one representative from each group. A paragraph-aware splitter chunks the remaining documents and copies parent metadata onto each chunk.

These filters reduce the candidate pool by over 95\% before chunk validation. We use Mistral-Large-3-675B \citep{mistral} as a chunk-level validator to remove false positives and assign petroleum-engineering subdomain labels. These labels support taxonomy-level coverage checks. A two-step majority-vote gate retains chunks that are informative, self-contained, and answerable, then assigns retained chunks to a thirteen-field petroleum-engineering taxonomy (Table~\ref{tab:domains}). We retain vote distributions to keep low-agreement
decisions inspectable.

\paragraph{Corpus Statistics.}
Table~\ref{tab:scale} summarizes the scale of the curated corpus and generated supervision. Every corpus row carries a stable chunk ID, source dataset, license
note, taxonomy label, and the decision trail of the filters above, allowing release review to include or exclude whole source families without re-running
the pipeline.

\begin{table}[t]
\centering
\small
\setlength{\tabcolsep}{3pt}
\begin{tabular}{@{}lr@{}}
\toprule
\multicolumn{2}{@{}l}{\textit{Curated corpus (\S\ref{subsec:curation})}} \\
\quad Chunks & 1{,}362{,}847 \\
\quad Characters & $\approx$8.0B \\
\quad Avg.\ tokens per chunk & $\approx$1{,}467 \\
\quad Token equivalents & $\approx$2.0B \\
\quad Energy classifier test accuracy & 98.39\% \\
\quad Energy classifier recall (energy) & 99.69\% \\
\midrule
\multicolumn{2}{@{}l}{\textit{Training snapshots (\S\ref{subsec:supervision})}} \\
\quad LLM-written negative rows (anchors) & $\approx$722k ($\approx$268k) \\
\quad Retrieval-mined hard-negative rows & $\approx$640k \\
\quad Embedding training rows (anchors) & $\approx$860k ($\approx$225k) \\
\quad Reranker candidate rows (pool / final) & $\approx$400k / $\approx$377k \\
\midrule
\multicolumn{2}{@{}l}{\textit{Filtered public release}} \\
\quad Retrieval triples: LLM / mined & $\approx$1.045M / $\approx$627k \\
\quad Reranker rows: LLM / RAG+LLM & $\approx$711k / $\approx$1.337M \\
\quad Strict-failure reranker rows & $\approx$190k \\
\bottomrule
\end{tabular}
\caption{Pipeline scale: curated corpus and generated retrieval supervision.}
\label{tab:scale}
\end{table}

\subsection{Synthetic Retrieval Supervision}
\label{subsec:supervision}
The curated corpus provides in-domain data rather than labeled relevance pairs; we therefore synthesize query-passage supervision for both embedding retrieval and reranking tasks.

Following Promptagator \citep{dai2022promptagator} and GPL \citep{wang2022gpl}, we first generate chunk-grounded queries and filter out unsupported or vague items; the accepted queries then anchor LLM-written hard negatives \citep{thakur2025hardnegatives} and candidate mining from deployed RAG system. The resulting records are packaged as contrastive triples for the embedding model and teacher-scored candidate rows for the reranker.

\paragraph{Query Generation.}
An instruction-tuned LLM generates three item types per chunk, matching the
query styles observed in user testing: natural-language questions, fact
statements, and terse keyword-style search queries. An item-level LLM filter
then rejects candidates that are unsupported by the chunk, too vague to act
as retrieval queries, or unanswerable from the chunk alone. The anchors that pass through this gate
feed both hard-negative branches. This filter removes 2.4\% of low-quality generated items.

\paragraph{Hard Negatives.}
For each anchor, we prompt an LLM with the query and its positive chunk to generate passages that are topically close but contain logical or semantic errors; a
validation pass rejects negatives that duplicate or accidentally answer the
query. We design the generator to cover entity substitutions, closely related but incorrect answers, and passages that preserve context while omitting the key value.

\paragraph{Reranking Data Generation.}
To create a reranking dataset, we deploy a baseline RAG stack (hybrid sparse+dense retrieval, then reranking) to generate adversarial candidate lists: we run generated queries against the fully indexed corpus and assign each case to a preserved bucket if the target chunk is found at rank 1 (Appendix~\ref{app:gen}), or to one of four failure buckets by rank.
Then, we re-label each retrieved candidate with a zero-shot LLM ranker Qwen3.6-35B-A3B. Such distillation is an effective strategy for creating efficient ranking models \citep{wu2025harnessing}. Each mined retrieval result and generated hard negative is given a teacher relevance score, using the ground-truth chunk as a reference point. To ensure that our training dataset is diverse, we balance sampling
across failure buckets, query-length buckets, and source IDs when assembling the candidate lists used for reranker training.

\paragraph{Training Data Construction.}
The final generation stage converts the accepted and mined records into two
training datasets. For the embedding model, each row is a triple
$(q, c, c_{\mathrm{neg},i})$, where $q$ is the generated query, $c$ is its
source chunk, and $c_{\mathrm{neg},i}$ is a hard negative. For the reranker,
each row is $(q, c'_i, y_i)$, where $c'_i$ is a retrieved or generated
candidate and $y_i$ is the teacher relevance score. During generation, we
deliberately cover only a fraction of the curated corpus: we cap generation
once in-domain validation gains plateau, with the LLM-negative export drawing
from about 2\% of the curated corpus. The full generation configuration is
reported in Appendix~\ref{app:gen} (Table~\ref{tab:app-gen}).

\section{Model Training}
\label{sec:training}
This section describes how PETRA's synthetic supervision (\S\ref{sec:dataset})
is used to adapt the two models of the deployed retrieval stack: the
first-stage embedding encoder and the cross-encoder reranker based on Qwen3-Embedding-8B and Qwen3-Reranker-8B models \citep{qwen2025embedding}. We adapt each
with lightweight LoRA adapters \citep{lora} over all linear layers with frozen
base weights, then manage the base-versus-domain tradeoff at inference time
with score fusion (first stage) and adapter merging (both stages).
Adapter-only training keeps checkpoints small, makes base-model upgrades and
rollbacks cheap, and bounds the release surface to weights demonstrably trained
on reviewed data.

\subsection{Adaptation}
\label{subsec:adaptation}

\paragraph{Embedding Adapter.}
Let $s_\theta(q,c)$ denote the query-chunk embedding similarity. The adapter
minimizes the standard InfoNCE contrastive loss
\citep{oord2018cpc,karpukhin2020dpr,wang2022e5}
\begin{equation}
\mathcal{L}_{\mathrm{emb}}
= -\log
\frac{e^{s_\theta(q,c)/\tau}}
{e^{s_\theta(q,c)/\tau} + \!\!\sum\limits_{c^-\in\mathcal{N}(q)}\!\! e^{s_\theta(q,c^-)/\tau}},
\end{equation}
where $\tau$ is the softmax temperature and $\mathcal{N}(q)$ contains the training dataset negatives, supplemented by in-batch negatives.

\paragraph{Reranker Adapter.}
The Qwen3 reranker scores a query--candidate pair $(q,c)$ by the logit
difference between its \emph{yes} and \emph{no} output tokens,
$r_\theta(q,c) = z^{\mathrm{yes}}_\theta(q,c) - z^{\mathrm{no}}_\theta(q,c)$,
where $z^{\mathrm{yes}}_\theta$ and $z^{\mathrm{no}}_\theta$ are the model's
output logits for the \emph{yes} and \emph{no} tokens. The pointwise yes/no
formulation offers a strong efficiency--effectiveness operating point for LLM
rerankers \citep{peng-etal-2025-efficiency}.
The adapter is trained pointwise with binary cross-entropy against labels
$y_i$, hard $1/0$ for source and negative chunks, and teacher scores for distilled rows:
\begin{equation}
\begin{split}
\mathcal{L}_{\mathrm{rr}}
= {} & -y_i \log \sigma(r_\theta(q,c'_i)) \\
     & -(1-y_i)\log\bigl(1-\sigma(r_\theta(q,c'_i))\bigr).
\end{split}
\end{equation}
The loss is pointwise, but rows are sampled from baseline top-$k$ candidate
lists containing both failures and preserve cases, so the reranker trains on
the same candidate distribution it scores at inference time.

\paragraph{Adapter Merging.}
To balance domain specialization and general retrieval performance in a single checkpoint, we merge adapters with TIES \citep{yadav2023ties}. For embeddings, we first train a
second LoRA adapter on public MS MARCO passage-ranking data
\citep{nguyen2016msmarco} as a general-retrieval anchor, then merge it with the energy adapter. For the reranker, we merge the earlier 50k-row pilot adapter with the final checkpoint, trading some specialization for the pilot's broader retention. During merging algorithm ablation, TIES outperformed the DARE-TIES \citep{yu2024dare} and equal-weight linear baseline.

\paragraph{Score Fusion.}
In deployment, we prioritize retrieval quality over encoder cost: the first stage runs both the base and adapted encoders (LoRA adapter, not a merged checkpoint) and combines their scores at inference time.

\vspace{-5mm}
\begin{equation}
s_{\mathrm{fuse}}(q,c)
= (1-\alpha)\, s_{\mathrm{base}}(q,c) + \alpha\, s_{\mathrm{adapter}}(q,c),
\end{equation}
with a single fixed weight $\alpha$ served for all traffic. We empirically find that this setup allows us to control the tradeoff between in-domain and out-of-domain performance at the cost of a second encoder pass. $\alpha$ was selected on a held-out dataset.

\subsection{Training Setup}
\label{subsec:training}

Both retrieval stages are adapted with LoRA \citep{lora} ($r{=}16$,
$\alpha{=}32$, dropout $0.05$) on all linear layers with frozen base
weights. The retrieval stage uses a learning rate of $5 \times 10^{-5}$, while the reranking stage uses $1 \times 10^{-5}$. The embedding adapter optimizes InfoNCE loss with the
Sentence-Transformers cached multiple-negatives ranking implementation
\citep{reimers2019sentencebert}, where in-batch negatives supplement explicit ones. The first stage serves a fixed fusion weight $\alpha{=}0.7$, selected in the in-domain held-out validation. Reranker teacher scores are produced by Qwen3.6-35B-A3B, which grades each retrieved candidate list on a
0--100 scale (three-rollout consensus, normalized to $[0,1]$). The reranker is trained in two stages: first on a balanced 50k-row pilot set, then on a 377k-row final set initialized from the pilot checkpoint. We use a constant schedule with 0.05 warmup.
The reranker TIES merge combines this final checkpoint with the earlier pilot adapter, and the embedding TIES merge uses an anchor adapter trained on the MS MARCO dataset \citep{nguyen2016msmarco}. 

\section{Experiments}
\label{sec:experiments}

\subsection{Evaluation Datasets}
\label{subsec:eval-datasets}

We evaluate on the operator's internal benchmark and fixed panel of public retrieval tasks designed to test both in-domain specialization and out-of-domain retention; per-task
sizes are in Appendix~\ref{app:benchmark-sizes}.

\textbf{SOP} is the operator's in-domain benchmark, built from internal
standard-operating-procedure documents and used as our primary measure of deployment
value; its assets remain internal, so we report aggregate scores only.
For public in-domain evaluation, we use Earth Science, a geoscience dataset from the
reasoning-intensive retrieval suite of \citet{su2025bright}, as the closest
public proxy for petroleum-engineering retrieval. \emph{Reasoning panel.} Six tasks from the same suite
(Earth Science, Robotics, AOPS, Economics, Sustainable Living, and
StackOverflow) probe reasoning-intensive retrieval across domains; we report
their macro average. For out-of-domain, we use SciFact \citep{wadden2020scifact},
FiQA \citep{maia2018fiqa}, and NFCorpus \citep{boteva2016nfcorpus} under the BEIR protocol \citep{thakur2021beir}, report their macro average, and use the same three tasks for both retrieval stages.

\subsection{Experimental Setup}

We report \ndcg \citep{jarvelin2002cumulated} as the primary metric, and
additionally recall@1 for first-stage retrieval, which matters operationally before
reranking. All numbers are single runs without seed variance. Training settings
are in \S\ref{subsec:training}. We include extensive evaluation tables in Appendix \ref{app:eval}. To position our stack against external systems,
we additionally compare against a leading publicly available retriever on the
BRIGHT leaderboard \citep{su2025bright,infretriever2025pro} on the tasks most
indicative of our deployment expectations (the operator SOP benchmark and Earth
Science), with the full comparison in Appendix~\ref{app:inf-x}.

\subsection{First-Stage Retrieval}

\begin{table}[t]
\centering
\small
\setlength{\tabcolsep}{4pt}
\begin{tabular}{@{}lccc@{}}
\toprule
First stage & SOP R@1 & SOP & OOD (3) \\
\midrule
Base (Qwen3-Emb.-8B) & 0.436 & 0.703 & \best{0.603} \\
Energy-only adapter & \second{0.495} & \second{0.757} & 0.498 \\
Fusion (fixed $\alpha{=}0.7$) & \best{0.503} & \best{0.763} & 0.570 \\
TIES merge (+MS MARCO) & 0.491 & 0.749 & \second{0.580} \\
\bottomrule
\end{tabular}
\caption{First-stage retrieval (\ndcg; R@1 = recall@1). OOD (3): macro over
SciFact, FiQA, and NFCorpus. Fusion uses a fixed $\alpha{=}0.7$ selected on
in-domain validation.}
\label{tab:embedding}
\vspace{-3mm}
\end{table}
Table~\ref{tab:embedding} shows that energy-only adapter delivers the in-domain
gain ($+0.054$ \ndcg, $+0.059$ recall@1) but costs $0.105$ \ndcg out of
domain; domain adaptation hurts general retrieval. Score fusion at fixed weight leads in domain (0.763 SOP, 0.503 recall@1) while keeping the base encoder in the loop, so it retains out-of-domain quality (0.570, versus 0.498 for the adapter and 0.603 for base).
The TIES energy+MS~MARCO merge is the best single-checkpoint compromise
(0.749 SOP, 0.580 out of domain, versus 0.536 and 0.567 for two linear
merges), recovering 0.082 of the energy-only adapter's out-of-domain loss but trailing the base.

\subsection{Reranking}

\begin{table}[t]
\centering
\small
\setlength{\tabcolsep}{3pt}
\begin{tabular}{@{}lcccc@{}}
\toprule
Reranker & Earth Sci. & Reason.\ (6) & SOP & OOD (3) \\
\midrule
Base reranker & 0.302 & 0.219 & 0.802 & 0.594 \\
Final checkpoint & \best{0.436} & \best{0.269} & \second{0.807} & \second{0.606} \\
TIES merge & \second{0.389} & \second{0.248} & \best{0.816} & \best{0.612} \\
\bottomrule
\end{tabular}
\caption{Reranking (\ndcg). Earth Sci.: public in-domain task; Reason.\ (6):
six-task panel; OOD (3): three-task panel. The final
checkpoint favors public in-domain transfer, while TIES favors SOP and OOD
retention.}
\label{tab:reranker}
\vspace{-3mm}
\end{table}

Table~\ref{tab:reranker} shows that candidate-list training improves every reported
surface over the base reranker. The final checkpoint is strongest on the
in-domain side: Earth Science rises from 0.302 to 0.436 ($+44\%$ relative)
and the reasoning panel from 0.219 to 0.269 ($+23\%$), while holding SOP and
the out-of-domain panel slightly above base reranker. The TIES merge shifts the
tradeoff toward retention, with the best SOP (0.816, $+1.7\%$) and out-of-domain score (0.612,
$+3.0\%$) at reduced Earth Science and reasoning-panel gains. No checkpoint
dominates; we keep both adapters and select by workload.

\subsection{Failed Recipes}

In early experiments, pairwise margin-ranking on retrieval-mined triples regressed on SOP and out-of-domain probes, and pointwise BCE on isolated LLM-written or retrieval-mined negative pairs reached high train-holdout accuracy that did not transfer to retrieval quality. We kept the native yes/no scoring path and pointwise BCE in the promoted recipe, but changed the training data labels to teacher-scored candidate rows balanced across failure and preserve cases. Retrieval-mined data was useful, but not as standalone hard-negative triples, but only once repackaged into these teacher-scored candidate lists.
Appendix~\ref{app:ablations} records the full ablation matrix.

\section{Conclusion}
\label{sec:conclusion}

Refinement of openly available web text (three public sources curated into a
1.36M-chunk corpus, about 2\% converted into synthetic supervision) was
enough to push a deployed retrieval stack past its strong base models in
domain while preserving general retrieval. Our work shows that aggressive data curation and quality control, knowledge distillation from strong teachers, and creating a small, focused dataset can create strong in-domain improvement.

\section*{Limitations}
The primary in-domain benchmark (SOP) is built from a single operator's
proprietary documents; we report aggregate scores only, so external
reproduction of the operator-side claims is not possible. The public panel is
fully reproducible, but it is a study-specific selection of nine tasks drawn
from public retrieval benchmarks, chosen to represent one public in-domain
task and a set of reasoning and out-of-domain tasks; we make no claims about
any full benchmark suite or leaderboard. All reported numbers are single
training and evaluation runs without seed-variance estimates; deltas on the
order of 0.01--0.02 \ndcg should be read with corresponding caution,
whereas the headline reranker gains ($+0.133$ Earth Science, $+0.050$ on the
reasoning panel) are larger.

Score fusion doubles first-stage encoder compute, and the deployed weight is
selected on in-domain validation only; we have not characterized the
latency/quality frontier or studied per-request fusion-weight selection,
which the validation sweeps suggest has clear headroom. The corpus and models
are English-only and centered on petroleum engineering; the energy classifier
and chunk gates were trained on labels distilled from commercial LLMs rather
than human annotation and inherit those models' biases and error modes.
Finally, the system is in user-testing deployment rather than full
production: evidence is
offline benchmarks plus operator validation, and no online A/B measurements
are available yet.

\section*{Ethical Considerations}
\paragraph{Proprietary data and privacy.}
The operator's documents and the SOP benchmark are proprietary. No document
text, queries, or per-query results are released; the paper reports aggregate
metrics only. Generated training rows derived from internal content are
excluded from any public release via the source-split manifests described in
the paper.

\paragraph{Web-derived data and licensing.}
The corpus is built from public datasets with recorded license families
(e.g.\ ODC-BY). A safety classifier screens unsafe content during curation,
and source manifests allow excluding source families whose redistribution
rights are unclear from any released artifact.

\paragraph{Safety-critical use.}
Retrieval errors in industrial documentation can surface wrong or outdated
procedures. The system is deployed for user testing as a human-in-the-loop
retrieval assistant rather than an autonomous decision system: it surfaces
source procedures for operators instead of acting on them. Continued monitoring of retrieval
quality on operator workloads is part of the deployment plan.

\paragraph{Domain scope.}
The corpus and models center on fossil-fuel engineering. The intended use is
achieving higher quality retrieval results for agentic and Q\&A workflows in oil and gas domain; the corpus's topical skew means coverage of
adjacent energy-transition topics is comparatively thin.

\bibliography{references}

\appendix
\section{Curation Operating Points}
\label{app:curation}

\begin{table}[ht]
\centering
\small
\begin{tabular}{@{}ll@{}}
\toprule
Filter & Threshold \\
\midrule
Document length & $100 \le$ words $\le 100{,}000$ \\
Non-alphanumeric ratio & $\le 0.30$ \\
Repeated-line fraction & $\le 0.70$ \\
Repeated-paragraph fraction & $\le 0.70$ \\
English (fastText) confidence & $\ge 0.30$ \\
Semantic-dedup similarity & $> 0.93$ removed \\
Chunk size / overlap / minimum & 2{,}048 / 100 / 100 tokens \\
LLM gate votes & $n=5$, plurality \\
\bottomrule
\end{tabular}
\caption{Heuristic and operating-point summary for the curation pipeline.}
\label{tab:app-heuristics}
\end{table}

\paragraph{Energy Classifier.}
The classifier fine-tunes Llama-3.1-8B with LoRA ($r{=}16$, $\alpha{=}32$,
dropout $0.05$; 45M trainable parameters, $0.56\%$ of the base) on 95{,}602
balanced documents (47{,}801 energy / 47{,}801 non-energy), split
76{,}480 / 9{,}560 / 9{,}562 for train/validation/test. Training used four
A100 80GB GPUs, bfloat16, effective batch size 64, learning rate
$2\times10^{-5}$ with cosine schedule and 10\% warmup; early stopping
selected step 1{,}100 after roughly two hours.
Table~\ref{tab:app-classifier}
summarizes performance on the validation and held-out test sets. The classifier
achieves high accuracy and F1 score while maintaining very high recall for
energy documents, which is the most important operating point for corpus
curation because false negatives remove potentially useful in-domain documents.

\begin{table}[t]
\centering
\small
\begin{tabular}{@{}lcc@{}}
\toprule
Metric & Validation & Test \\
\midrule
Accuracy & 98.55\% & 98.39\% \\
F1 & 98.56\% & 98.41\% \\
Precision & 97.54\% & 97.17\% \\
Recall & 99.60\% & 99.69\% \\
ROC-AUC & 99.76\% & 99.76\% \\
\bottomrule
\end{tabular}
\caption{Energy classifier performance.}
\label{tab:app-classifier}
\end{table}

To further characterize classifier behavior, Table~\ref{tab:app-confusion}
reports the confusion matrix on the held-out test set. The results show that
most classification errors are false positives (non-energy documents predicted
as energy), while false negatives are rare, consistent with the classifier's
high recall objective.

\begin{table}[t]
\centering
\small
\begin{tabular}{@{}lcc@{}}
\toprule
 & Pred.\ non-energy & Pred.\ energy \\
\midrule
Actual non-energy & 4{,}642 & 139 \\
Actual energy & 15 & 4{,}766 \\
\bottomrule
\end{tabular}
\caption{Test-set confusion matrix (9{,}562 documents; 154 errors, 1.61\%).}
\label{tab:app-confusion}
\end{table}

\begin{figure*}[t]
    \centering
    \includegraphics[width=\textwidth]{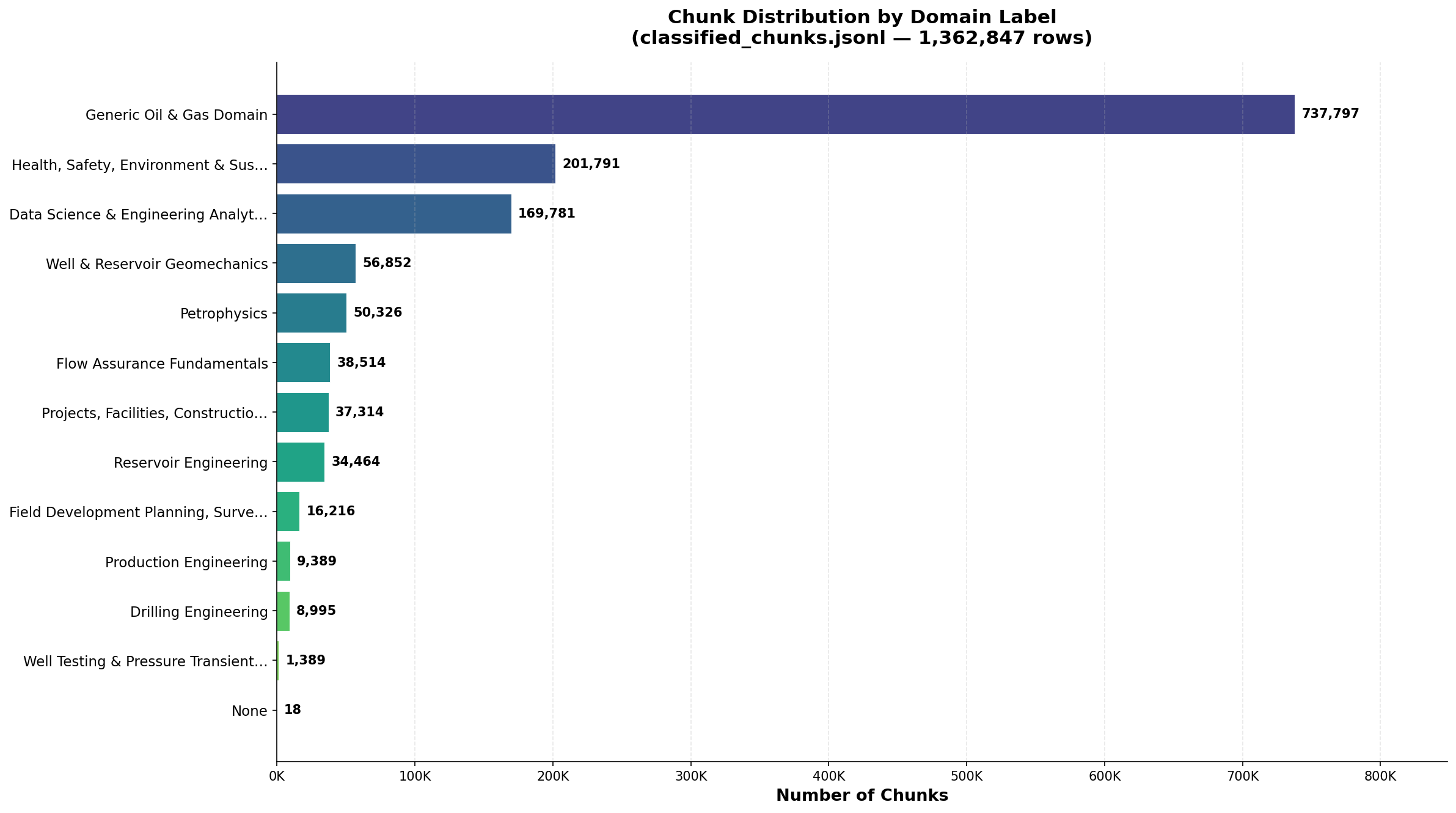}
    \caption{Distribution of curated energy chunks across the oil-and-gas domain taxonomy.}
    \label{fig:class}
\end{figure*}

\paragraph{Oil-and-gas Domain Taxonomy.}
Qualified energy chunks are assigned to one of thirteen oil-and-gas domain
categories. Table~\ref{tab:domains} defines the taxonomy used for labeling,
while Figure~\ref{fig:class} shows the resulting distribution of curated chunks
across these categories.

\begin{table*}[t]
\centering
\small
\renewcommand{\arraystretch}{1.18}
\begin{tabular}{@{}c p{4.4cm} p{8.2cm}@{}}
\toprule
\textbf{\#} & \textbf{Category} & \textbf{Scope (abridged)} \\
\midrule
1  & Generic Oil \& Gas Domain & General industry content not specific to one sub-discipline (overviews, terminology, history, markets, general operations). \\
2  & Health, Safety, Environment \& Sustainability & HSE, safety management, environmental impact, emissions, sustainability, regulations, risk. \\
3  & Data Science \& Engineering Analytics & Data analytics, machine learning, statistics, and digital/computational methods applied to oil \& gas. \\
4  & Well \& Reservoir Geomechanics & Poroelasticity, in-situ stress, wellbore stability, compaction/subsidence. \\
5  & Petrophysics & Porosity/permeability, core analysis, well logging, formation evaluation. \\
6  & Flow Assurance Fundamentals & Multiphase flow, hydrates/wax/asphaltene/scale, thermal management. \\
7  & Projects, Facilities, Construction \& Operations & Surface facilities, infrastructure, projects, construction, operations. \\
8  & Reservoir Engineering & PVT, flow in porous media, material balance, recovery (incl.\ EOR/CO$_2$). \\
9  & Field Development Planning, Surveillance \& Economics & Volumetrics, reserves (PRMS), forecasting (DCA), NPV/CAPEX/OPEX, surveillance. \\
10 & Production Engineering & IPR, lift (gas-lift/ESP/rod/PCP), surface processing, nodal analysis. \\
11 & Drilling Engineering & Well design, drilling fluids/hydraulics, well control, completions/stimulation. \\
12 & Well Testing \& Pressure Transient Analysis & Flow regimes, test interpretation, $k$/$s$/$C$/boundary estimation. \\
13 & None & Does not fit any category above. \\
\bottomrule
\end{tabular}
\caption{The thirteen-category oil-and-gas domain taxonomy used for domain labelling by
the document-level filter of Section~\ref{sec:chunk}.}
\label{tab:domains}
\end{table*}

\section{Dataset Construction Details}
\label{app:gen}

Table~\ref{tab:app-gen} gives the full configuration of the synthetic
supervision generated for PETRA (\S\ref{subsec:supervision}); aggregate
scales are in Table~\ref{tab:scale}.

\begin{table}[t]
\centering
\small
\begin{tabular}{@{}p{0.30\linewidth}p{0.62\linewidth}@{}}
\toprule
Stage & Configuration \\
\midrule
Query items per chunk & Three styles: natural-language question, fact statement, keyword-style search query. \\
Query filter & Reject items unsupported by the chunk, too vague to retrieve, or unanswerable from the chunk alone. \\
LLM hard negatives & Generator sees query and positive; validation rejects negatives that duplicate or accidentally answer the query; failure modes varied (wrong facility, neighboring question, key fact omitted). \\
Retrieval-mined buckets & By rank of the known positive: preserve (rank 1), Fail@1 (2--10), Fail@10 (11--30), Fail@30 (31--50), Fail@50 (absent). \\
Strict-failure export & Hardest peS2o cases expanded into $\approx$190k public-release rows. \\
Candidate-list sampling & Balanced across failure buckets, query-length buckets, and source IDs. \\
\bottomrule
\end{tabular}
\caption{Synthetic generation configuration for PETRA.}
\label{tab:app-gen}
\end{table}

\paragraph{Training-row Assembly.}
The accepted items are normalized into two training surfaces. The LLM-written
negative export contributes $\approx$722k validated rows, and retrieval-mined
hard negatives add $\approx$640k rows whose distractors are real inference-time
candidates rather than synthetic text. These are assembled into $\approx$860k
embedding triples over $\approx$225k anchors and a $\approx$400k-row
reranker candidate pool carrying baseline ranks, selection buckets, and teacher
scores; the final reranker trains on the $\approx$377k rows that remain after
removing problems shared with the 50k-row warm-start stage 
(Table~\ref{tab:scale}). The filtered public release retains only FinePDFs,
peS2o, and Wikipedia rows: $\approx$1.045M LLM-negative retrieval triples,
$\approx$627k retrieval-mined triples, $\approx$711k LLM-only reranker rows,
$\approx$1.337M combined RAG+LLM reranker rows, and $\approx$190k peS2o
strict-failure reranker rows. PETRA deliberately covers only a fraction of the
corpus: we cap generation once in-domain validation gains are sufficient (the
LLM-only reranker export draws on $\approx$27k distinct chunks, about 2\% of the
corpus) rather than exhaustively converting all 1.36M chunks.

\section{Evaluation Benchmark Sizes}
\label{app:benchmark-sizes}

Table~\ref{tab:benchmark-sizes} lists the retrieval benchmarks used in
\S\ref{sec:experiments}. Query counts represent  evaluated queries. Document counts
are the corresponding retrieval corpus sizes. The SOP benchmark uses a
row-self-match protocol over full internal contexts; its counts are reported,
but the underlying text and per-query results are proprietary.

\begin{table*}[t]
\centering
\small
\setlength{\tabcolsep}{5pt}
\begin{tabular}{@{}llrr@{}}
\toprule
Panel & Benchmark & Queries & Documents \\
\midrule
Internal in-domain & SOP & 3{,}607 & 3{,}607 \\
\midrule
Public in-domain / reasoning & BRIGHT Earth Science & 116 & 121{,}249 \\
Reasoning & BRIGHT Robotics & 101 & 61{,}961 \\
Reasoning & BRIGHT AOPS & 111 & 188{,}002 \\
Reasoning & BRIGHT Economics & 103 & 50{,}220 \\
Reasoning & BRIGHT Sustainable Living & 108 & 60{,}792 \\
Reasoning & BRIGHT StackOverflow & 117 & 107{,}081 \\
\midrule
Out-of-domain & SciFact & 300 & 5{,}183 \\
Out-of-domain & FiQA & 648 & 57{,}638 \\
Out-of-domain & NFCorpus & 323 & 3{,}633 \\
\bottomrule
\end{tabular}
\caption{Evaluation benchmark sizes. SOP counts refer to the internal
row-self-match evaluation set; public counts are for the evaluated public tasks
used in Table~\ref{tab:embedding} and Table~\ref{tab:reranker}.}
\label{tab:benchmark-sizes}
\end{table*}

\section{External Retriever Comparison}
\label{app:inf-x}

To situate our adapted stack against a strong publicly available baseline, we
compare with \texttt{inf-retriever-v1-pro} \citep{infretriever2025pro}, a
leading publicly available dense retriever on the BRIGHT leaderboard
\citep{su2025bright}, evaluated under our protocols on the operator SOP
benchmark and on Earth Science. We run it on its own and paired with its
companion query rewriter (\texttt{inf-query-aligner}), which distills complex
prompts into concise search queries. On the deployment distribution (SOP), our
adapted stack leads by a wide margin: our fused first stage reaches 0.763 \ndcg
and our reranker 0.807, versus 0.688 for the external retriever, whose aligner
regresses it further to 0.614. The aligner is, however,
query-distribution-specific: on the BRIGHT-style reasoning queries of Earth
Science it lifts the external retriever sharply, from 0.417 to 0.722 \ndcg,
while it hurts the operator's already-concise SOP queries. Since SOP queries
already match the concise form the retriever expects, the aligned configuration
is not the operating point relevant to our deployment; a fuller characterization
of this model across query-rewriting regimes is left to future work.

\begin{table}[t]
\centering
\small
\setlength{\tabcolsep}{6pt}
\begin{tabular}{@{}lcc@{}}
\toprule
System & SOP & Earth Sci. \\
\midrule
Ours (promoted stack) & \best{0.807} & 0.436 \\
\texttt{inf-retriever-v1-pro} & 0.688 & 0.417 \\
\quad + \texttt{inf-query-aligner} & 0.614 & \best{0.722} \\
\bottomrule
\end{tabular}
\caption{External-baseline comparison (\ndcg; best per column in bold). Our
deployed stack leads on the operator SOP distribution; the external query
aligner helps the BRIGHT-style reasoning queries of Earth Science but regresses
the operator's concise SOP queries.}
\label{tab:inf-x}
\end{table}

\section{Reranker Training Ablations}
\label{app:ablations}

The reranker was developed through a sequence of ablations spanning supervision strategies, training objectives, and adapter-merging methods. Table~\ref{tab:app-reranker-ablations} summarizes the principal experiments and their outcomes. Retrieval-mined pairwise training and isolated pointwise supervision did not produce the strongest downstream results, whereas candidate-row pointwise distillation consistently improved benchmark performance and was adopted in the final recipe. Additional experiments explored retention-oriented adapters and model-merging approaches, with TIES providing the most favorable retention–adaptation tradeoff among the tested variants.

\begin{table}[t]
\centering
\small
\setlength{\tabcolsep}{3.5pt}
\begin{tabular}{@{}p{0.30\linewidth}p{0.40\linewidth}p{0.19\linewidth}@{}}
\toprule
Ablation & Observed outcome & Decision \\
\midrule
Retrieval-mined-only pairwise (margin-ranking on mined triples) & Broad regression on SOP and out-of-domain probes. & Reject. \\
Isolated-negative pointwise (BCE on LLM-written or mined pairs) & High train-holdout accuracy; unreliable downstream transfer. & Diagnostic only. \\
Candidate-row pointwise distillation (failure/preserve rows, teacher scores) & Improved Earth Science and the reasoning panel over base. & Promote (final). \\
Retention-oriented adapter (smaller balanced run) & Stronger retention, weaker adaptation transfer. & Merge input. \\
Larger failure/preserve continuation & Best Earth Science / reasoning panel; some SOP and out-of-domain tradeoff. & Final checkpoint. \\
TIES merge (retention adapter + final) & Recovered retention, kept part of adaptation gain. & Tradeoff variant. \\
DARE-TIES variants & Weaker than TIES in tested runs. & Not promoted. \\
Linear-average reranker merge & Checked; lower than TIES on SOP, BEIR-3, and BRIGHT-3. & Reject. \\
Teacher scoring format & Listwise graded scoring cheaper than pairwise-reference with no clear quality loss. & Use listwise. \\
\bottomrule
\end{tabular}
\caption{Reranker ablation families behind the final recipe.}
\label{tab:app-reranker-ablations}
\end{table}

\section{Hard-Negative Example}
\label{app:example}

Table~\ref{tab:hardneg} illustrates the two hard-negative patterns used during training. The first negative contains factually correct information but refers to the wrong facility, while the second remains grounded in the correct facility but answers a related question rather than the one being asked. Both negatives remain lexically similar to the positive passage, forcing the reranker to rely on entity identity and supporting evidence rather than surface-level term overlap.

\begin{table}[t]
\centering
\small
\renewcommand{\arraystretch}{1.2}
\begin{tabular}{@{}p{0.21\columnwidth} p{0.73\columnwidth}@{}}
\toprule
\textbf{Query} & ``Which processing units were added to the refinery, and what did the 2007 hydrotreater and 2009 hydrogen unit enable?'' \\
\midrule
\textbf{Positive} & \emph{Refinery~A} added a fluid catalytic cracker and alkylation unit (1980s), a distillate hydrotreater (2007), and a hydrogen unit (2009), enabling ultra-low-sulfur diesel and greater crude flexibility. \\
\addlinespace
\textbf{Negative\newline (wrong facility)} & \emph{Refinery~B} added a distillate hydrotreater (2007) and a hydrogen unit (2009), enabling ultra-low-sulfur diesel and greater crude flexibility. \\
\addlinespace
\textbf{Negative\newline (wrong evidence)} & Refinery~A upgraded \emph{emissions monitoring and wastewater treatment} in the same period, omitting the hydrotreater and hydrogen-unit effects the query asks for. \\
\bottomrule
\end{tabular}
\caption{A query with its positive chunk and two abridged hard negatives.
Italics mark the discriminating content: each negative copies the positive's
terminology but changes the facility (top) or the evidence (bottom).}
\label{tab:hardneg}
\end{table}

\section{Benchmark Contamination Scan}
\label{app:contamination}

Because several evaluation benchmarks are derived from publicly available sources, some degree of document overlap with the training corpus is possible. We therefore performed a contamination scan to quantify the extent of overlap between the training data and the evaluation benchmarks. Using both exact matching of normalized text and partial matching based on 16-token shingles, we searched for benchmark documents that also appeared in the training corpus. Table~\ref{tab:gold-docs-by-surface} reports the number of overlapping gold documents identified in each benchmark.
Although a small number of overlapping documents were found, we did not observe overlap in benchmark queries or in the positive and negative passages used for evaluation. To measure the impact of these overlaps, we repeated the evaluation after removing all affected examples. The resulting metrics, shown in Table~\ref{tab:leakage-filtered-reranker}, differ only marginally from the original scores, suggesting that document overlap has negligible impact on the reported evaluation results.

\begin{table}[t]
\centering
\small
\begin{tabular}{@{}p{0.70\columnwidth}r@{}}
\toprule
Evaluation dataset & Overlap \\
\midrule
SciFact / FiQA / NFCorpus / AOPS / Economics / Robotics & 0 \\
Earth Science & 17 \\
StackOverflow & 3 \\
Sustainable Living & 3 \\
SOP & 2 \\
\bottomrule
\end{tabular}
\caption{Number of gold documents that overlap with the training set.}
\label{tab:gold-docs-by-surface}
\end{table}

\begin{table*}[t]
\centering
\footnotesize
\setlength{\tabcolsep}{4pt}
\begin{tabular}{@{}lrrrrr@{}}
\toprule
System variant & SOP & Earth Sci. & Sust. & StackOverflow & BRIGHT-6 \\
\midrule
Base retrieval
  & 0.707 (-0.0001)
  & 0.342 (+0.0079)
  & 0.179 (+0.0033)
  & 0.179 (-0.0113)
  & 0.188 (-0.0000) \\
Base reranker
  & 0.801 (-0.0001)
  & 0.311 (+0.0083)
  & 0.218 (+0.0040)
  & 0.222 (-0.0118)
  & 0.219 (+0.0001) \\
Final checkpoint
  & 0.807 (-0.0001)
  &\textbf{ 0.445} (+0.0096)
  & \textbf{0.252} (+0.0019)
  & \textbf{0.325} (-0.0099)
  & \textbf{0.269} (+0.0003) \\
TIES merge
  & \textbf{0.816} (-0.0001)
  & 0.396 (+0.0063)
  & 0.219 (+0.0024)
  & 0.287 (-0.0098)
  & 0.248 (-0.0002) \\
\bottomrule
\end{tabular}
\caption{Leakage-filtered reranker results (\ndcg; best per column in bold).
Values are filtered scores; parentheses show absolute change from the original
metric after excluding affected gold-document-overlap evaluation queries.}
\label{tab:leakage-filtered-reranker}
\end{table*}

\section{Extended Evaluation Tables}
\label{app:eval}
This section reports the complete evaluation results underlying the summary metrics presented in the main paper. Table~\ref{tab:first-stage-adaptation-tradeoff} examines the tradeoff between in-domain adaptation and out-of-domain retention for the first-stage retriever, while Table~\ref{tab:reranker-domain-transfer-retention} reports the corresponding transfer and retention behavior of the reranker checkpoints.

\begin{table*}[t]
\centering
\footnotesize
\setlength{\tabcolsep}{4pt}
\begin{tabular}{@{}lrrrrrrrrrrr@{}}
\toprule
System variant & SOP & Earth Sci. & Earth $\Delta$ & StackOverflow & Econ. & Sust. & Robotics & AOPS & SciFact & FiQA & NFCorpus \\
\midrule
Base retrieval & 0.703 & 0.334 & -- & 0.190 & 0.178 & 0.176 & 0.151 & \best{0.099} & 0.795 & 0.613 & 0.405 \\
Base reranker & 0.802 & 0.302 & -- & 0.233 & 0.235 & 0.214 & 0.256 & 0.072 & 0.801 & 0.555 & 0.425 \\
Final checkpoint & 0.807 & \best{0.436} & \best{+44\%} & \best{0.335} & \best{0.261} & \best{0.250} & \best{0.262} & 0.070 & 0.811 & 0.574 & 0.433 \\
TIES merge & \best{0.816} & 0.389 & +29\% & 0.297 & 0.248 & 0.216 & 0.258 & 0.081 & \best{0.817} & \best{0.584} & \best{0.436} \\
\bottomrule
\end{tabular}
\caption{Reranker domain-transfer and retention (\ndcg; best per column in bold).}
\label{tab:reranker-domain-transfer-retention}
\end{table*}

\begin{table*}[t]
\centering
\small
\setlength{\tabcolsep}{6pt}
\begin{tabular}{@{}lrrrrr p{4.6cm}@{}}
\toprule
First-stage variant & SOP & SOP R@1 & $\Delta$ SOP & OOD (3) & $\Delta$ OOD & Readout \\
\midrule
Base retrieval & 0.703 & 0.436 & -- & \best{0.603} & -- & strongest OOD anchor \\
Energy LoRA & 0.757 & 0.495 & +0.054 & 0.498 & -0.105 & strong in-domain gain, clear forgetting \\
Score fusion, $\alpha=0.7$ & \best{0.763} & \best{0.503} & \best{+0.060} & 0.570 & -0.033 & best SOP, much less forgetting \\
TIES merge (+MS MARCO) & 0.749 & 0.491 & +0.046 & 0.580 & -0.023 & best single-checkpoint retention \\
\bottomrule
\end{tabular}
\caption{First-stage adaptation tradeoff (\ndcg; best per column in bold).}
\label{tab:first-stage-adaptation-tradeoff}
\end{table*}
\section{Training and Inference Pipeline Implementation}
\label{app:system-pipeline}

Training and inference is implemented as a set of batch-oriented pipelines. The data-generation
pipeline starts from chunked corpus rows and produces positive anchors,
filtered query variants, LLM-written hard negatives, retrieval-mined hard
negatives, and final training datasets. The embedding-training pipeline
converts these rows into InfoNCE triples for the first-stage encoder. The
reranker-training pipeline starts from baseline top-$k$ candidate lists, attaches
teacher scores, and trains the cross-encoder on candidate rows that match the
distribution scored at inference time.

Corpus-scale generation runs under Slurm job orchestrator with Ray Data as the row-level
orchestrator. Slurm allocates GPU nodes, while Ray partitions the 
dataset rows into map tasks. Inference is done via HTTP using external \texttt{vLLM} servers. Each allocated node
starts a \texttt{vLLM} endpoint inside the Slurm allocation; Ray workers send
batches to the discovered endpoints after automated health checks pass. Smaller models use
one GPU per server, while larger MoE models use tensor parallel serving across
the GPUs on a node. This separates cluster scheduling, row-level parallelism,
and model-serving throughput.

Training uses PyTorch, Transformers, PEFT, Sentence-Transformers, and Hugging
Face Accelerate. The first-stage embedding adapter is trained as a LoRA adapter
over frozen base weights with a cached multiple-negatives InfoNCE objective,
bf16 mixed precision, gradient checkpointing, gradient accumulation, and
distributed data-parallel launches through Accelerate. The reranker uses the
same adapter-only design, but is trained with Accelerate/FSDP: transformer-layer
wrapping, bf16 mixed precision, full parameter/gradient/optimizer sharding,
CPU-efficient model loading, and sharded checkpoint state. We use FSDP rather
than full-model replication because the cross-encoder has substantially higher
memory pressure than the embedding objective.

Evaluation follows the same batch-oriented design. To speed up large volume of evaluations, first-stage retrieval can be
cached as text embeddings and ranked candidate arrays. Reranker evaluations then
load candidate IDs from the cache and run only the cross-encoder stage. Cache
keys are semantic: they include query and document identifiers, text
fingerprints, query formatting, retrieval model identity, and fusion settings,
but exclude runtime-only choices such as GPU ID, batch size, attention backend,
Slurm job ID, and worker count. This lets checkpoint matrices reuse expensive
retrieval work while keeping the compared retrieval protocol fixed.

\end{document}